\documentstyle[11pt,newpasp,twoside,epsf]{article}
\def\plotone#1{\centering \leavevmode\epsfxsize=\columnwidth \epsfbox{#1}}

\begin{document}

\title{H$\alpha$ emission line profiles of selected post-AGB stars} 
\author{Griet C. Van de Steene \& Peter R. Wood}
\affil{RSAA (ANU), Mount Stromlo Observatory, Private Bag, Weston Creek,
ACT 2611, Australia}
\author{Peter A.M. van Hoof}
\affil{CITA, McLennan Labs, 60 St. George Street, Toronto, ON M5S 3H8, Canada }

\begin{abstract}

We present H$\alpha$ emission line spectra of 7 post-AGB stars.
Typically, they have a P-Cygni type profile with a very strong
emission component and a relatively narrow absorption component. These
lines are formed in the fast post-AGB wind which in most cases has a
velocity of around 100~km/s. Hence the emission originates close to
the central star and the absorption occurs in the fast wind itself or
in the region where the fast wind sweeps up the slow wind.  The
emission profile has very broad wings.  It is still unclear what
causes these wings.

\end{abstract}

\section{Introduction}

A most intriguing challenge is to understand how Asymptotic Giant
Branch (AGB) stars transform their surrounding mass-loss shells in a
couple of thousand years into the variety of shapes and sizes observed
in Planetary Nebulae (PNe).  There are a number of theories currently
being investigated. In the generalized interacting stellar wind 
model, a variety of axisymmetric PN shapes are obtained by the
interaction of a very fast central star wind with the progenitor AGB
circumstellar envelope (Kwok 1982), when the latter is denser near the
equator than the poles (Frank et al.\ 1993). Sahai \& Trauger (1998)
proposed that the primary agent for shaping PNe is not the density
contrast, but a high speed collimated outflow of a few 100 km/s.
No consensus about the dominant physical process responsible for the
shaping of PNe has emerged so far, but there is agreement that they
occur during the early AGB-to-PN transition stage.  However, the
details of the rates of evolution and the strength of the stellar
winds during the AGB-to-PN transition phase are very poorly known,
either theoretically or observationally.  To remedy this we have
started to study a sample of post-AGB objects spectroscopically. 

\section{Observations}

Objects with far infrared colors typical of PNe were selected
from the IRAS catalogue. Apart from PNe, only post-AGB stars are
typically found in this part of the color-color diagram.  Furthermore,
the objects selected were not detected in the radio continuum above a
detection limit of 3 mJy (Van de Steene \& Pottasch 1993). Hence they
don't seem to have evolved to the PN stage as yet.

We obtained JHKL images of the candidates with CASPIR on the 2.3-m
telescope at Siding Spring Observatory in Australia in order to obtain
accurate positions for the IRAS counterparts.  Subsequently, high
resolution optical spectra were obtained with the Double Beam
Spectrograph on the 2.3-m telescope. The resolution was 0.55~\AA\ per
pixel.

We present H$\alpha$ profiles of selected post-AGB stars showing H$\alpha$
in emission.  Obviously not all objects showed H$\alpha$ in emission,
some have H$\alpha$ in absorption and some show neither H$\alpha$
emission or absorption. Some of the post-AGB stars detected in the
near infrared could not be seen in the optical due to large
extinction.

\section{H$\alpha$ emission line profiles}

\begin{figure}[t]
\plotone{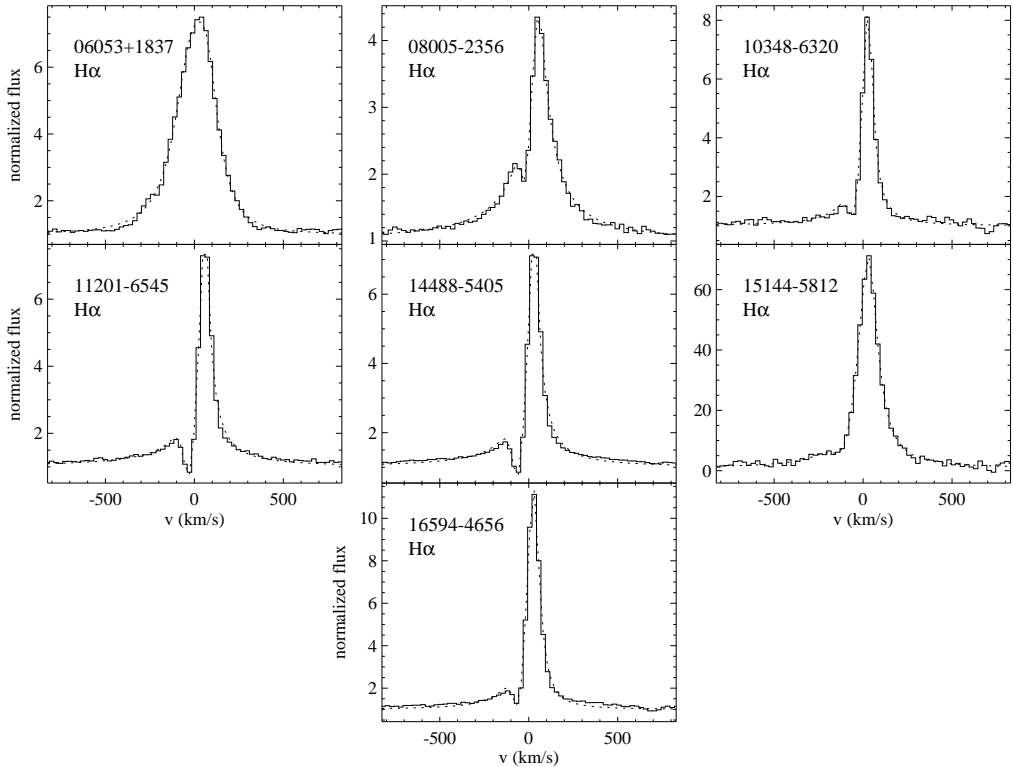 }
\caption{ 
Plots of the H$\alpha$ profiles of the post-AGB stars (solid lines)
with the fits over-plotted (dotted lines).}
\end{figure}

The wings are not fitted well when only gaussian profiles are used.
We fitted the H$\alpha$ emission lines with the sum of a `generalized'
Lorentz emission profile and a gaussian absorption profile. A
`generalized' Lorentz profile is defined as $F_\lambda
\propto 1/[1 + (\Delta\lambda/\lambda_{\rm w})^\alpha]$ with 
$\lambda_{\rm w}$ (the HWHM of the profile) and $\alpha$ as free
parameters. For $\alpha = 2$ this is the familiar Lorentz profile.
Since all observed profiles are wider on the blue side, the HWHM of
the fitted profile was allowed to have different values in the red and
the blue part (across the absorption).  The results of the fits are
shown in Table~\ref{results}.  The listed absolute flux is for the
whole, integrated profile and has not been dereddened.  The listed
FWHM is the sum of the blue and red HWHM.  The entries $v_{\rm abs}$
and $v_{\rm emm}$ are the velocities of the center of the absorption
and the emission peak relative to the systemic velocity.  We
determined the systemic velocity from photospheric absorption lines
using C\,{\sc ii} at 6578.05~\AA\ and 6582.88~\AA\, and He\,{\sc i} at
6678.15~\AA, if detected.  We used the equivalent widths of these
lines to determine the spectral types of the objects by comparing them
with the results of Lennon et al.\ (1993).  We calculated the terminal
velocity of the wind $v_{\infty}$ from the blue edge of the absorption
in the H$\alpha$ profile according to: 
$v_{\infty}$ = FWHM$_{\rm abs}-v_{\rm abs}$.

\begin{table}
\caption{Fit parameters for the H$\alpha$ emission lines.}
\label{results}
\begin{tabular}{llrrrrrr}
\hline
Object & Spec. & Flux & FWHM & $\alpha$ & $v_{\rm abs}$ & $v_{\rm emm}$ & $v_{\infty}$ \\
IRAS   & Type & $10^{-16}$ W/m$^2$ & km/s & & km/s & km/s & km/s \\
\hline
06053$+$1837 &      &    3.13 & 253 & 2.63 &   --- & --- & --- \\
08005$-$2356 &      &    8.38 & 149 & 1.45 & $-16$ &  57 &  72 \\
10348$-$6320 & B4   &    0.78 &  77 & 1.78 & $-48$ &  30 & 114 \\
11201$-$6545 & A0   &    5.63 &  75 & 1.45 & $-24$ &  61 &  99 \\
14488$-$5405 & B9Ia &   19.15 &  79 & 1.49 & $-55$ &  31 & 131 \\
15144$-$5812 &      &    0.41 & 135 & 1.98 &   --- & --- & --- \\
16594$-$4656 & B7   &    2.55 &  93 & 1.99 & $-49$ &  28 & 126 \\
\hline
\end{tabular}
\end{table}

\section{Interpretation}

The similarity in the H$\alpha$ profiles leads us to suggest the
following interpretation of our observations.

Trammel et al.\ (1994) determined that the H$\alpha$ emission of IRAS
08005$-$2356 was polarized.  Based on the polarization angles, they
suggested a bipolar geometry for this object.  A bipolar geometry was
also suggested for IRAS 16594$-$4656 based on HST images by Hrivnak et
al.\ (1999). Based on the similarity of the presented emission line
profiles, quite a few of our objects could be bipolar PNe.

The lack of other forbidden emission lines in the spectra indicates
that the H$\alpha$ emission is not due to photo-ionization of the
shell.  IRAS 16594$-$4656 is the only star in our sample for which [O\,{\sc
i}] emission was observed. No [N\,{\sc ii}] or [S\,{\sc ii}] emission
was detected in any of the spectra.  Consequently the main H$\alpha$
emission cannot be caused by shock emission either.  The blue peak in
the H$\alpha$ emission is too faint compared to the red peak to
originate from a disk or expanding shell.

The P-Cygni type of the H$\alpha$ profiles indicates the presence of a
strong stellar wind which can produce the emission line and the
blue-shifted absorption.  The terminal velocity of the wind can be
derived from the position of the blue absorption edge. The velocity of
the fast wind is around 100~km/s. No value for $v_\infty$ could be
determined for IRAS 06053$+$1837 and IRAS 15144$-$5812, but
indications are that they have a higher velocity than the rest of the
sample.

The absorption is too deep to be photospheric, hence the absorption
must be produced outside the emission region. For post-AGB stars the
fast stellar wind is encircled by the slower AGB wind, which can
absorb line photons produced by the fast wind. The velocity of the
center of the absorption is similar to the AGB wind velocity or higher.
This suggests that the absorption occurs in the fast wind itself or
in the region where the fast wind sweeps up the slow wind.

The H$\alpha$ profiles show very extended wings, which we can follow
in our spectra up to $\sim$500 km/s and even 1500~km/s for IRAS
08005$-$2356.  In the case of HD101584, a type B9II post-AGB star with
a spectrum very similar to our sample, Bakker et al.\ (1996)
attributed the wings to electron scattering by free electrons in the
stellar wind.  However, Thomson scattering is wavelength independent
and the fact that we don't see wings in our Br$\gamma$ spectra argues
against this.  Multiple Rayleigh scattering off neutral hydrogen in
the expanding AGB shell or multiple scattering off dust grains in that
same shell may be able to produce these wings as well.  
However, Rayleigh scattering has too low a cross section to be of much
importance in a dusty environment like a post-AGB star.  Grain
scattering is supported by Trammel et al.\ (1994) who suggested that
the polarization of H$\alpha$ was most likely caused by scattering off
dust grains in a dense torus surrounding the central star in IRAS
08005$-$2356.  However, for both cases, the fact that the observed
velocities in the wings are so much larger than the AGB or post-AGB
wind velocities casts doubt on this explanation. To reach the observed 
velocities the scattering must occur in the wind or maybe in a Keplerian
disk, if there is one. It is still unclear what causes them. 

Extended wings have also been observed in young compact PNe such as
M2$-$9 (Balick 1989), Mz$-$3 (L\'opez \& Meaburn 1983), Cn~3$-$1 and
M3$-$27 (Miranda et al.\ 1997). No profound explanation for these
wings has been proposed in the literature.

\end{document}